\documentclass[a4paper,11pt]{article}
\pdfoutput=1

\usepackage{amssymb}
\usepackage{amsmath}
\usepackage{commath}
\usepackage{bm}

\usepackage[numbers,square,sort&compress]{natbib}

\usepackage{graphicx}
\usepackage{color}
\usepackage{caption}
\usepackage{subcaption}

\usepackage{authblk}

\usepackage{tikz}
\usetikzlibrary{calc}

\usepackage{titlesec}
\titleformat{\section}
  {\normalfont\large\bfseries\uppercase}{\thesection}{1em}{}
\titleformat{\subsection}
  {\normalfont\normalsize\bfseries}{\thesubsection}{1em}{}
\titleformat{\subsubsection}
  {\normalfont\normalsize\itshape}{\thesubsubsection}{1em}{}
\titleformat{\paragraph}[runin]
  {\normalfont\normalsize\itshape}{\theparagraph}{1em}{}

\newcommand{\new}[1]{#1} 

\newcommand{\xb}{\bm{x}}
\newcommand{\kb}{\bm{k}}
\newcommand{\Mb}{\bm{M}}
\newcommand{\nb}{\bm{n}}
\newcommand{\eb}{\bm{e}}

\renewcommand{\eqref}[1]{Eqn~(\ref{#1})}
\renewcommand\cite{\citep}

\usepackage{hyperref}

\title{An acoustic space-time and the Lorentz transformation in aeroacoustics}
\date{\today}
\author[1,3,*]{Alastair L Gregory}
\author[2,**]{Samuel Sinayoko}
\author[1,$\dagger$]{Anurag Agarwal}
\author[1,$\ddagger$]{Joan Lasenby}
\affil[1]{Department of Engineering,
University of Cambridge,
Trumpington Street,
Cambridge, CB2 1PZ}
\affil[2]{Institute of Sound and Vibration Research,
University of Southampton,
Highfield,
Southampton, SO17 1BJ}
\affil[3]{Magdalene College,
University of Cambridge,
Cambridge,
CB3 0AG}
\affil[*]{\texttt{alg57@cam.ac.uk} (author for correspondence)}
\affil[**]{\texttt{s.sinayoko@soton.ac.uk}}
\affil[$\dagger$]{\texttt{aa406@cam.ac.uk}}
\affil[$\ddagger$]{\texttt{jl221@cam.ac.uk}}
\begin{document}
\maketitle
\begin{abstract}
  In this paper we introduce concepts from relativity and geometric
  algebra to aeroacoustics. We do this using an acoustic space-time
  transformation within the framework of sound propagation in uniform
  flows. By using Geometric Algebra we are able to provide a simple
  geometric interpretation to the space-time transformation, and are
  able to give neat and lucid derivations of the free-field Green's
  function for the convected wave equation and the Doppler shift for a
  stationary observer and a source in uniform rectilinear motion in a
  uniform flow.

\end{abstract}

\section{Introduction}
This is a curiosity driven study that aims to explore the
applicability of concepts from relativity and geometric algebra in the
field of aeroacoustics. The analogy of acoustics with relativity has
been investigated in physics and cosmology, but less has been done to
use this work in the field of aeroacoustics. Geometric algebra has
been successfully applied to a variety of fields but aeroacoustics has
not yet been one of these. In order to introduce these concepts we
make use of the simple problem of sound propagation in a uniform flow.
Even though this problem is well understood, we show that by using the
new concepts presented in this paper, we obtain a new geometric
interpretation of the transformation that relates sound propagation in
a uniform flow to the no flow case.

\subsection{Analogies with Relativity}
There is a substantial body of work in physics and cosmology on the
analogies between general relativity and noise propagation in fluid
flows. Much of it is fundamental and originates from the seminal work
of Unruh \cite{Unruh:1981ue} who identified the analogy between black
holes and noise propagation in supersonic flows. Several other
analogies exist between the relativistic physics of black holes and
cosmology and various engineering disciplines, including optics and
electrical engineering \cite{Visser:2013tr}. These analogies have
resulted in the creation of several analogue space-times. The use of
these analogues has so far been largely restricted to physicists
attempting to gain a better understanding of gravity and quantum
mechanics \cite{Visser:1998wd,Barcelo:2004vg}. This has recently
changed with the recognition that engineers can benefit from the
insights and techniques developed by physicists to deepen their own
understanding and develop new technologies.  Applying general
relativity to optics has given birth to the field of Transformation
Optics~\cite{Leonhardt:2009ui,Leonhardt:2006gt} which has led to the
development of perfect lenses and electromagnetic
cloaks~\cite{Leonhardt:2006gt} using metamaterials, and has since
stimulated similar studies on acoustic
cloaks~\cite{Zhang:2011cj,Pendry:2008eb,Torrent:2008im}.  Relativity
has also been applied successfully in electrical engineering
\cite{Pendry:2006hq,Leonhardt:2006jl}.

In this paper we introduce an analogy between relativity and
aeroacoustics within the context of sound propagation in a uniform
flow. We make extensive use of an acoustic space-time that provides a
geometric interpretation of the transformation approach to solving the
convected wave equation. We demonstrate the geometry established by a
uniform flow field and use this to explain the form of the Green's
function for the convected wave equation \cite{Garrick:1953txa}, as
well as the appearance of Doppler shifts
\cite{Bateman:1917tv,Gerkema:2013hi} and convective amplification.

\subsection{Geometric Algebra}
Geometric Algebra (GA) makes it easier to draw an analogy between
relativity and aeroacoustics. GA was pioneered by
Hestenes~\cite{Hestenes:2002tv,Hestenes:1992vw} in the 1960s; this
work took the algebras of Clifford and Grassmann
\cite{Clifford:1878cw,Grassmann:1844wx} and developed them into a
powerful geometric language for mathematics and physics. Since the
1960s GA has been used in many physics and engineering
applications. The key feature of GA is its ability to express physical
equations in a coordinate-free form; the algebra is also extendible to
any dimension. In this way GA is able to subsume complex numbers,
quaternions, tensors etc. and provide simple and intuitive geometric
interpretations of objects and often operators. GA has been applied
successfully in electromagnetism
\cite{Doran:2003jd,Arthur:2011ty,Abbott:2010wy}, where the electric
and magnetic fields are written as one geometric quantity and the four
Maxwell's equations are then reduced to a single equation. It has also
been used for analysis of conformal arrays in radar and sonar
applications \cite{Zou:2011cg}. Given the analogies between acoustics
and electromagnetism \cite{Jones:1989wu,Carcione:1995ea}, it might
seem reasonable to suppose that aeroacoustics might also benefit from
a GA approach. While GA has been used in information engineering
\cite{Dorst:2002ti} and mechanical engineering \cite{McRobie:1999wv}
there have been few attempts at applying it in fluid mechanics. This
paper is the first attempt at applying the language of GA to
aeroacoustics. While much of the analysis for this paper was derived
in GA, the reader will only explicitly need GA in
\secref{sec:GeometricAlgebra}. Other GA functionality which is
required is given in the Appendices.

\subsection{Sound In a Uniform Flow}
In order to make use of the concepts introduced above, we consider the
classic problem of sound propagation in a uniform flow. The equivalent
problem of analysing the sound from the moving source was studied in
the context of electromagnetics in the early 20th century, for example
see Stratton~\cite{Stratton:1941wc}, who gave the resulting signal
amplitude distributions relative to the source either at time of
emission or reception. Lighthill later gave the sound distributions in
the context of acoustics \cite[\S4.1]{Lighthill:1962jp}. Lighthill
effectively solved for the sound field in the frame of the fluid, then
used a Galilean transformation to find the solution in the frame of
the observer in which flow is present
\cite[Eq.~(16)]{Lighthill:1962jp}. Lighthill also gave some physical
interpretation of the result in the fluid frame. A clear treatment of
this approach is given by Dowling and Williams
\cite[\S9]{Dowling:1983vo}.

More recently, the problem has been solved by considering the
convected wave equation, and using a Lorentz-type transform
\cite[\S9.1.1]{Rienstra:2013vd}. This is useful because the
non-convected wave equation is easily solved using an appropriate
Green's function. This latter approach has been used to develop a wide
range of aeroacoustic theories in active control~\cite{Joseph:1998tn},
duct acoustics~\cite{Chapman:1994gl,Sinayoko:2010cz}, aerofoil
broadband noise~\cite{Garrick:1953txa,Amiet:1976fh,Roger:2005kr}, and
aerodynamic theories on thin aerofoils in compressible
flows~\cite{Kussner:1940uz,Graham:1970fj}. The transforms used in
these methods have been summarised by \cite{Chapman:2000ky}.

However, the physical interpretation of this transformation approach
is not clear, with the transform itself generally simply stated in
terms of similarity variables with little justification as to where
the transformation originates from and how it can be generalised to
other problems. For example, it is not clear how the theory of
Blandeau et al~\cite{Blandeau:2011ib} on trailing edge noise for
rotating blades, which neglects the uniform mean flow effects, can be
extended to take these effects into account; Sinayoko et
al~\cite{Sinayoko:2013iy} therefore resorted to
solving the same problem from first principles without the use of
transformations.

Addressing this shortcoming, and providing an interpretation of the
transformation method is the main goal of this paper. We shall see
that this method differs from that of Lighthill and Dowling et al in
that the solution is found not in the frame of the fluid, but in a
third frame that moves with the observer \emph{and} where the wave
operator takes a simple form. The advantage over Lighthill's method is
that boundary conditions that are given in the observer frame can be
more easily applied in the new third frame, which we shall refer to as
the Lorentzian frame because this is the type of transform required to
produce it.

\subsection{Organisation of the Paper}
The paper is organised as follows. \secref{sec:TheTransform} defines
the \emph{acoustic space-time} and introduces the observer, fluid and
Lorentzian frames as well as the associated time-space coordinates and
transformations.  \secref{sec:TheTransform-frequency} provides the
reciprocal frequency-wavenumber coordinates and transformations.
These transformations are used to derive simple solutions for three
classic aeroacoustic problems in \secref{sec:examples}: the free field
Green's function for the convected wave equation in a uniform flow;
the Doppler shift for a stationary observer and a source in uniform
rectilinear motion in a uniform flow; \new{providing a comparison with
  previous work and examples of the advantages of our approach.}
\secref{sec:GeometricAlgebra} introduces the language of Geometric
Algebra and shows how it provides new insights into the geometry
underlying the transformations and the nature of the wave operator.

\section{Transforming the convected wave equation in space-time coordinates}
\label{sec:TheTransform}
Let $(x,y,z)$ be Cartesian coordinates in a Euclidean space, and
$p(x,y,z,t)$ be a scalar function of position and time $t$ that
represents the perturbation to the pressure in a fluid occupying the
space. If the background pressure and density of the fluid are
constant, and the fluid is stationary, the perturbation pressure must
satisfy \cite{Pierce:1981vd},
\begin{equation}
  \label{eq:waveEq}
  \left[\frac{\partial^2}{\partial x^2} + \frac{\partial^2}{\partial y^2}
  + \frac{\partial^2}{\partial z^2} - \frac{1}{c^2}\frac{\partial^2}
  {\partial t^2}\right]p = -f,
\end{equation}
where $f(x,y,z,t)$ is a scalar function that represents the sound
sources in the fluid, and $c$ is the (constant) speed of sound in the
fluid. If instead there is a uniform background flow with velocity $U$
in the $z$ direction, then $p$ must satisfy,
\begin{equation}
  \label{eq:convectedWaveEq}
  \left[\frac{\partial^2}{\partial x^2} + \frac{\partial^2}{\partial y^2}
  + \frac{\partial^2}{\partial z^2} - \frac{1}{c^2}\left(\frac{\partial}
  {\partial t} + U\frac{\partial}{\partial z}\right)^2\right]p = -f.
\end{equation}

We will present and interpret a transformation that allows us to solve
\eqref{eq:convectedWaveEq} by transforming it into
\eqref{eq:waveEq}, solving that, and then transforming back. This
transformation has been presented most recently in an alternative form
by \cite{Chapman:2000ky} and has been used prior to this, for example in
\cite{Martinez:1980bf,Martinez:1983ci}, however, in previous presentations,
no interpretation of the transformation was given. We rectify this
using the concept of a newly defined ``acoustic space-time'' that is
analogous to the special relativistic space-time of physics, but with
the speed of light replaced by the speed of sound.

\new{It will be immediately obvious to readers familiar with
  relativity that the wave equation is Lorentz invariant, but that the
  physical transformation between \eqref{eq:waveEq} and
  \eqref{eq:convectedWaveEq} is a Galilean transform, since the
  underlying physics is Newtonian. Usually this is regarded as a
  curiosity, however by making use of the Lorentz symmetry in a
  hyperbolic space we end up with a problem that is easier to solve.
  On top of this we can then construct non-orthogonal frames using the
  Galilean transformation to obtain solutions in terms of physical
  variables. Having introduced a hyperbolic background space it is
  natural to use Geometric Algebra, in which rotors provide a lucid
  way of handling Lorentz transformations, and the whole problem can
  be written in a coordinate free form.

In the upcoming section the Galilean and Lorentz transformations are
introduced in detail within an acoustic space-time. Readers already
familiar with this can move on to the description of how the
space-time can be used to interpret transformations in the
frequency-wavenumber domain \secref{sec:TheTransform-frequency},
applications \secref{sec:examples}, and the advantages provided by
geometric algebra \secref{sec:GeometricAlgebra}.}

\subsection{The Acoustic Space-Time}
\new{We will be concerned throughout the paper with vector spaces with
dimension higher than 3, and more significantly, with mixed signature.
The 3-dimensional Euclidean space usually used to model Newtonian
mechanics can have an arbitrary set of basis vectors (a frame),
denoted $\{\eb_i\}$, which are defined relative to each other by the
set of inner products,}
\begin{equation}
  \eb_i\cdot\eb_j.
\end{equation}
\new{To deal with spaces of higher dimension we simply need more basis
vectors, but to deal with spaces of mixed signature we need to make
one further generalisation. In Euclidean space the inner products
$\eb_i\cdot\eb_j$ are all positive numbers. In a space of mixed
signature this restriction is relaxed. In this article we deal with a
4-dimensional space in which we define a frame $\{\gamma_\mu'\}$ that
satisfies the simple space-time metric,}
\begin{equation}
  \label{eq:spaceTimeSig}
  \gamma_\mu'\cdot\gamma_\nu' = \begin{bmatrix}
    1&0&0&0\\0&-1&0&0\\0&0&-1&0\\0&0&0&-1
  \end{bmatrix}
  \text{ where } \mu,\nu=t,x,y,z.
\end{equation}
\new{Note that we have not used bold font to denote these basis vectors. In
general we reserve bold font for vectors in 3-dimensional Euclidean
space. We have also switched from using Latin indices to using Greek
indices. We reserve Latin indices for frames spanning 3-dimensional
Euclidean space. This matrix of all the inner products of a frame is
called the metric of the frame. Signature refers to the signs of the
eigenvalues of the metric, which is a fundamental property of the
space. We can see from this metric that there are a mixture of signs
in the metric of space-time, hence the term mixed signature.}

For any arbitrary frame $\{e_\mu\}$ it is possible to construct a
reciprocal frame $\{e^\mu\}$ (see \appref{app:reciprocalFrames}) such
that,
\begin{equation}
  e_\mu\cdot e^\nu = \delta^\nu_\mu,\quad
  \delta^\nu_\mu = \left\{\begin{array}{ll}
    1 & \mu=\nu \\ 0 & \mu\neq\nu
  \end{array}\right. ,
\end{equation}
We use this reciprocal frame to find the coordinates of an arbitrary
vector when the original frame is not orthonormal, as in this
case. For our $\{\gamma_\mu'\}$ frame, the reciprocal frame
$\{{\gamma^\mu}'\}$ is given by,
\begin{equation}
  {\gamma^t}' = \gamma_t',\quad {\gamma^x}' = -\gamma_x',\quad
  {\gamma^y}' = -\gamma_y',\quad {\gamma^z}' = -\gamma_z'.
\end{equation}

An arbitrary four vector $\xi$ can be written in the $\{\gamma_\mu'\}$
frame as,
\begin{equation}
  \xi = ct'\gamma_t' + x'\gamma_x' + y'\gamma_y'
  + z'\gamma_z',
\end{equation}
and the coordinates can be found using the reciprocal frame,
\begin{equation}
  \label{eq:fluidFrameCoordDefn}
  ct' = \xi\cdot{\gamma^t}',\quad
  x' = \xi\cdot{\gamma^x}',\quad
  y' = \xi\cdot{\gamma^y}',\quad
  z' = \xi\cdot{\gamma^z}'.
\end{equation}
\new{In this article we only deal with frames that are the same at every
position in the vector space, and with spaces that are not curved. For
these kinds of space the method used above for finding coordinates
using the reciprocal frame works for any frame (see
\cite{Doran:2003jd} and \appref{app:reciprocalFrames}).} Note that the
time coordinate has been factored into $c$ and $t'$, where $c$ is the
speed of sound in the fluid being considered, and $t'$ has units of
time.

\subsubsection{Galilean Transform}
\label{sec:galileanTransform}
Let us now define a second frame $\{\gamma_\mu\}$, using the scalar
constant $M=U/c$, as,
\begin{equation}
  \label{eq:GalileanTransformDefn}
  \gamma_t = \gamma_t' - M\gamma_z',\quad
  \gamma_x = \gamma_x',\quad \gamma_y = \gamma_y',\quad
  \gamma_z = \gamma_z'.
\end{equation}
The reciprocal frame $\{\gamma^\mu\}$ is given by (see
\eqref{eq:findingRecipFrame}),

\begin{equation}
  \gamma^t = \gamma_t',\quad \gamma^x = -\gamma_x',\quad
  \gamma^y = -\gamma_y',\quad
  \gamma^z = M\gamma_t' - \gamma_z'.
\end{equation}

An arbitrary four vector $\xi$ can be written in the $\{\gamma_\mu\}$
frame as,
\begin{equation}
  \xi = ct\gamma_t + x\gamma_x + y\gamma_y
  + z\gamma_z,
\end{equation}
and the coordinates can be found using the reciprocal frame,
\begin{equation}
  \label{eq:observerFrameCoordDefn}
  ct = \xi\cdot{\gamma^t},\quad
  x = \xi\cdot{\gamma^x},\quad
  y = \xi\cdot{\gamma^y},\quad
  z = \xi\cdot{\gamma^z}.
\end{equation}
\new{From this point on we redefine $(x,y,z,t)$ as the coordinates of
  this frame in a four dimensional space-time defined below. We will
  see however that $(x,y,z,t)$ will still represent the space and time
  that are measurable by us, which is why we choose to retain these
  symbols.} Using these expressions for the coordinates along with
\eqref{eq:fluidFrameCoordDefn}, we can show that the following
coordinate transformations hold,
\begin{equation}
  \label{eq:galileanCoordTrans}
  ct = ct',\quad x = x',\quad y = y',\quad z = z' + Mct'.
\end{equation}
From \eqref{eq:galileanCoordTrans}, if an observer is at constant
$(x,y,z)$, then they are moving in the $-z$ direction with Mach number
$M$ with respect to the $(x',y',z')$ coordinates. We can say that
$\{\gamma_\mu'\}$ is the frame of the fluid, while $\{\gamma_\mu\}$ is
the frame of the observer. \new{It is clear that this is simply a Galilean
  transformation.}

Let us now define the linear operator $\mathcal L$ that operates on
fields in the acoustic space-time we have constructed, and write it in
terms of the coordinates in the $\{\gamma_\mu'\}$ frame,
\begin{equation}
  \mathcal L = \frac{\partial^2}{\partial{x'}^2} +
  \frac{\partial^2}{\partial{y'^2}} +
  \frac{\partial^2}{\partial{z'^2}} - \frac{1}{c^2}
  \frac{\partial^2}{\partial{t'^2}}.
\end{equation}
Using the chain rule and the coordinate transformations given in
\eqref{eq:galileanCoordTrans} we can show that $\mathcal L$ may be
written as,
\begin{equation}
  \label{eq:LinObserverFrame}
  \mathcal L = \frac{\partial^2}{\partial x^2} +
  \frac{\partial^2}{\partial y^2} +
  \frac{\partial^2}{\partial z^2} - \frac{1}{c^2}\left(
  \frac{\partial}{\partial t} + U\frac{\partial}{\partial z}
  \right)^2.
\end{equation}
If we consider the scalar functions $p(\xi)$ and $f(\xi)$, then the
equation,
\begin{equation}
  \label{eq:coordinateFreeWaveEq}
  \mathcal Lp = -f,
\end{equation}
is exactly equivalent to \eqref{eq:waveEq} if $\mathcal L$ is
written in the $\{\gamma_\mu'\}$ frame, and equivalent to
\eqref{eq:convectedWaveEq} if $\mathcal L$ is written in the
$\{\gamma_\mu\}$ frame. Note that $p$ and $f$ are scalar functions of
position in space-time, $\xi$. This four-vector can be parameterised
in terms of the coordinates of any frame we introduce, and hence $p$
(and $f$) can be thought of as $p(\xi(x,y,z,t))$, or
$p(\xi(x',y',z',t'))$ (we have omitted the factor of $c$ since it is a
constant). \new{In fact, $\mathcal L$ can be defined in a simple coordinate
free manner using the vector derivative of Geometric Algebra. In
addition, the use of Geometric Algebra for this purpose allows an
intuitive explanation to be given as to why $\mathcal L$ takes a
simple form in one frame, and a complex one in the other, as is
explained in \secref{sec:GAVectorDerivative}.}

We have found a transformation that allows us \new{to} rewrite
\eqref{eq:convectedWaveEq} as \eqref{eq:waveEq}, however, if a
source or observer is at rest in the $\{\gamma_\mu\}$ frame
(i.e. they are at a constant $x$, $y$ and $z$), they will be in motion
in the $\{\gamma_\mu'\}$ frame. We would like to transform to a
frame where the wave operator takes the simple form in
\eqref{eq:waveEq}, but where an observer or source that is
stationary in the $\{\gamma_\mu\}$ frame is still stationary.

\subsubsection{Lorentz Transform}
\label{sec:lorentzTransform}
Let us define the frame $\{\gamma_\mu''\}$ as,
\begin{equation}
  \label{eq:LorentzTransformDefn}
  \gamma_t'' = \frac{1}{\beta}(\gamma_t' - M\gamma_z'),\quad
  \gamma_x'' = \gamma_x',\quad \gamma_y'' = \gamma_y',\quad
  \gamma_z'' = \frac{1}{\beta}(\gamma_z' - M\gamma_t'),
\end{equation}
where $\beta$ is defined by,
\begin{equation}
  \beta = \sqrt{1 - M^2}.
\end{equation}
Note that it is simple to show that the $\{\gamma_\mu''\}$ frame
satisfies the space-time metric in \eqref{eq:spaceTimeSig}. The
reciprocal frame $\{{\gamma^\mu}''\}$ is given by (see
\eqref{eq:findingRecipFrame}),
\begin{equation}
  {\gamma^t}'' = \gamma_t'',\quad {\gamma^x}'' = -\gamma_x''
  ,\quad {\gamma^y}'' = -\gamma_y'',\quad
  {\gamma^z}'' = -\gamma_z''.
\end{equation}

An arbitrary four vector $\xi$ can be written in the
$\{\gamma_\mu''\}$ frame as,
\begin{equation}
  \xi = ct''\gamma_t'' + x''\gamma_x'' + y''\gamma_y'' +
  z''\gamma_z'',
\end{equation}
and the coordinates can be found using the reciprocal frame,
\begin{equation}
  ct'' = \xi\cdot{\gamma^t}'',\quad
  x'' = \xi\cdot{\gamma^x}'',\quad
  y'' = \xi\cdot{\gamma^y}'',\quad
  z'' = \xi\cdot{\gamma^z}''.
\end{equation}
Using these expressions for the coordinates along with
\eqref{eq:fluidFrameCoordDefn} and
\eqref{eq:observerFrameCoordDefn} we can derive the coordinate
relations,
\begin{equation}
\label{eq:prime-to-double-prime}
  ct'' = \frac{1}{\beta}(ct' + Mz'),\quad
  x'' = x',\quad y'' = y',\quad
  z'' = \frac{1}{\beta}(z' + Mct'),
\end{equation}
\begin{equation}
  \label{eq:nop_To_pp_coordTrans}
  ct'' = \beta ct + \frac{M}{\beta}z,\quad x'' = x,\quad y'' = y,\quad
  z'' = \frac{z}{\beta}.
\end{equation}
We note from the relations in \eqref{eq:nop_To_pp_coordTrans} that
if a source or observer is at constant $x,y,z$ they will also be at
constant $x'',y'',z''$. \new{Using the chain rule along with these
  relations, we can show that,}
\begin{equation}
  \label{eq:L_pp_coords}
  \mathcal L = \frac{\partial^2}{\partial{x''}^2} +
  \frac{\partial^2}{\partial{y''^2}} +
  \frac{\partial^2}{\partial{z''^2}} - \frac{1}{c^2}
  \frac{\partial^2}{\partial{t''^2}}.
\end{equation}
\new{We see that under the Lorentz transformation the wave equation is left
unchanged. This will be obvious to readers familiar with relativity,
and is particularly well explained in a coordinate free way using
Geometric Algebra. The Lorentz transform can be shown very simply to
be a generalised form of rotation \secref{sec:GALorentzTrans}, and
hence leaves the metric unchanged. Keeping the metric unchanged
automatically means that the operator $\mathcal L$, which can be
written in a frame independent way using Geometric Algebra, remains
simple, as is explained in \secref{sec:GAVectorDerivative}.}

\begin{figure}
  \centering
  \begin{tikzpicture}[scale=4.0]
    \def\M{0.5}
    \def\Monb{0.577}
    \def\oneonb{1.155}
    \draw[arrows=->,ultra thick,color=green] (0,0) -- (-\Monb,\oneonb)
         node[at end,right,font=\Large] {$\gamma_t''$};
    \draw[arrows=->,ultra thick,color=green] (0,0) -- (\oneonb,-\Monb)
         node[at end,above,font=\Large] {$\gamma_z''$};
    \draw[arrows=->,ultra thick] (0,0) -- (0,1)
         node[very near end,right,font=\Large] {$\gamma_t'$};
    \draw[arrows=->,ultra thick] (0,0) -- (1,0)
         node[very near end,above,font=\Large] {$\gamma_z'$};
    \draw[arrows=->,thick,color=red] (0,0) -- (-\M,1)
         node[very near end,left,font=\Large] {$\gamma_t$};
    \draw[arrows=->,thick,color=red] (0,0) -- (1,0)
         node[very near end,below,font=\Large] {$\gamma_z$};
    \draw[arrows=->,dashed,ultra thin] (0,1) -- (-\M,1);
    \node at (0.1,1.2) {$-M\gamma_z'$};
    \draw[arrows=->,ultra thin]
         (0.1,1.15) to[out=-90,in=90] (-0.2,1);
  \end{tikzpicture}
  \caption{The effect of the Galilean and Lorentz transforms on the
    basis vectors.}
  \label{fig:galileanAndLorentzTrans}
\end{figure}
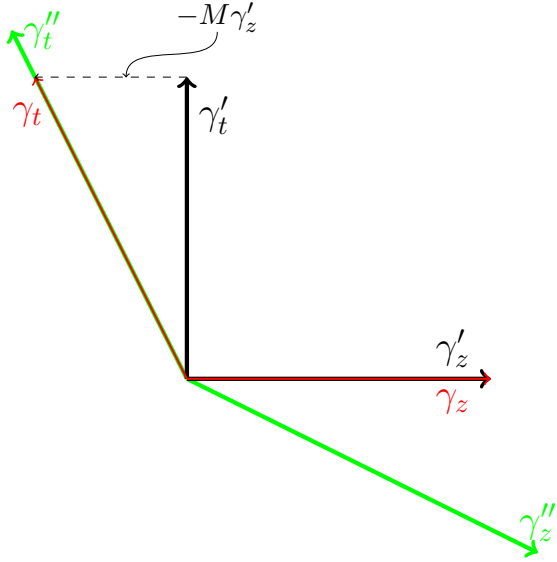

For now though let us consider these transformations more carefully.
Since $x$ and $y$ are unaffected, we can illustrate the transformation
by considering only the $z$ and time directions
(\figref{fig:galileanAndLorentzTrans}).

Conceptually we can think of our method as beginning by creating a
frame that moves with the fluid. This frame is the $\{\gamma_\mu'\}$
frame, and satisfies the space-time signature
(\eqref{eq:spaceTimeSig}). This is why in
\figref{fig:galileanAndLorentzTrans} this is the frame shown with
its basis vectors orthogonal. The speed of sound is fundamental to
this frame.

From this frame we then create two new frames, using either a Galilean
or a Lorentz transform. To create the space with the Galilean
transform we skew the time vector in the direction of the flow (this
is the red frame in \figref{fig:galileanAndLorentzTrans}), and
leave the spatial vectors unchanged. The coordinates of this frame
represent the coordinates that we observe if we sit still and see
fluid flowing past us. In this frame the complex wave operator given
in \eqref{eq:LinObserverFrame} applies.

The frame created using the Lorentz transform ($\{\gamma_\mu''\}$)
has its time vector pointing in the same direction as that of the
observer frame, but this frame must satisfy the space-time
signature. Hence the time vector is stretched and the spatial vector
must change (this is the green frame in
\figref{fig:galileanAndLorentzTrans}). It must be remembered when
looking at \figref{fig:galileanAndLorentzTrans} that we are
considering a hyperbolic space, so $\gamma_t''$ and $\gamma_z''$
\emph{are} orthogonal. In this frame the simple wave operator applies,
and we also have the important property that if an observer is at
constant $x,y,z$ they are also at constant $x'',y'',z''$. This allows
the conversion of boundary conditions from the $\{\gamma_\mu\}$
frame to the $\{\gamma_\mu''\}$ frame.

\subsection{\new{Transforming in the frequency-wavenumber domain}}
\label{sec:TheTransform-frequency}
The ordinary wave equation takes \new{a simpler algebraic form} in the
frequency-wavenumber domain,
\begin{equation}
  \label{eq:helmholtz}
  \left[\frac{\omega^{2}}{c^{2}} - k^2 \right] P(\bm k, \omega) =
  F(\bm k, \omega),
\end{equation}
where $P$ is the Fourier transform of $p$, defined as,
\begin{equation}
  \label{eq:fourier}
  P(\bm k, \omega) = \int_{-\infty}^{\infty}\int_{\mathbb R^3}
  {p(\bm x, t) e^{i(\omega t - \bm k \cdot \bm x)}} d^3\bm x dt
\end{equation}
and $F$ is the Fourier transform of $f$. \eqref{eq:helmholtz} can be
obtained by taking the Fourier transform of \eqref{eq:waveEq}.

Similarly, the convected \new{wave equation (\eqref{eq:convectedWaveEq})
  takes a simpler algebraic form in the frequency-wavenumber domain,}
\begin{equation}
  \label{eq:helmholtz-convected}
  \left[ \left(\frac{\omega}{c} - Mk_z\right)^2 - k^2 \right]
  P(\bm k, \omega) = F(\bm k, \omega).
\end{equation}

\eqref{eq:helmholtz-convected} can be converted into
\eqref{eq:helmholtz} by using an appropriate transformation. The main
difference with the physical domain transformations is that, given a
reference frame $\{\gamma_{\mu}\}$, the wavenumbers $k_x$, $k_y$ and
$k_z$ and the frequency $\omega$ are the components of a 4-wavevector
$\kappa$ defined in the reciprocal frame $\{\gamma^{\mu}\}$ as,
\begin{equation}
  \label{eq:kappa}
  \kappa \equiv \frac{\omega}{c}{\gamma^t} -
  k_x {\gamma^x} - k_y{\gamma^y} - k_z{\gamma^z},
\end{equation}
or, equivalently,
\begin{align}
  \label{eq:kappa-coordinates}
  \omega/c &\equiv \kappa \cdot \gamma_t,&
  k_x &\equiv -\kappa \cdot \gamma_x, &
  k_y &\equiv -\kappa \cdot \gamma_y, &
  k_z &\equiv -\kappa \cdot \gamma_z.
\end{align}
This definition of $\kappa$ is such that the phase
$\phi\equiv\kappa\cdot\xi$ takes the usual form,
\begin{equation}
  \label{eq:phase}
  \phi = \omega t - k_x x - k_y y - k_z z,
\end{equation}
irrespective of the reference frame.

Since we have already defined the fluid frame $\{\gamma_{\mu}'\}$ and
the Lorentzian frame $\{\gamma_{\mu}''\}$, we can readily obtain the
wavevector coordinate transformations by applying
\eqref{eq:kappa-coordinates} to the appropriate frame, which
yields,
\begin{align}
  \label{eq:freq-domain-fluid-coordinates}
  \omega' &= \omega - M c k_z, & k_z' &= k_z, & k_x' &= k_x,
  & k_y' &= k_y.
\end{align}
for the fluid frame, and,
\begin{align}
  \label{eq:freq_components_gppframe1}
  \omega'' &= \frac{1}{\beta}(\omega' + M c k_z') = \frac{\omega}{\beta} , \\
  k_z'' &= \frac{1}{\beta}\left(k_z' + M \frac{\omega'}{c} \right) =
  \beta k_z + \frac{M}{\beta} \frac{\omega}{c} , \\
  k_x'' &= k_x' = k_x, \\
  k_y'' &= k_y' = k_y.
\end{align}
for the Lorentzian frame.

Thus, from \eqref{eq:freq-domain-fluid-coordinates}, the frequency
depends on the axial wavenumber in the fluid frame
($\omega'=\omega+Mck_z'$), while the wavenumber components remain the
same in the observer and the fluid frames ($k_z=k_z'$, $k_x=k_x'$ and
$k_y=k_y'$). This behaviour is the reverse of the way the physical
coordinates transform between the observer and the fluid frames: time
remains the same ($t=t'$) while the axial coordinate changes
($z'=z-Mct'$).

Similarly, in the Lorentzian frame, the frequency $\omega''$ is fixed
relative to the observer frequency $\omega$ while the axial wavenumber
$k_z''$ is a linear combination of $k_z$ and $\omega$. In contrast, in
the space-time domain, it is the axial Lorentzian position $z''$ that
is fixed relative to the axial observer position $z$ ($z''=z/\beta$)
while the Lorentzian time $t''$ is a linear combination of $z$ and
$t$.

\section{Applications}
\label{sec:examples}
\subsection{Sound Propogation in Acoustic Space-Time Along Null Directions}
\label{sec:nullPropogation}
Let us imagine a point marker on a wave front moving at the speed of
sound $c$ relative to the fluid, and in any spatial direction. Written
using the coordinates of the frame that moves with the fluid
$\{\gamma_\mu'\}$, the marker's motion will be described by,
\begin{equation}
  \label{eq:soundParticleMotion}
  \sqrt{(x'-x_s')^2+(y'-y_s')^2+(z'-z_s')^2} = c(t'-\tau'),
\end{equation}
if at time $\tau'$ the point is at the position defined by
$x_s',y_s',z_s'$. It is important to note that the time of the frame
that moves with the fluid is the same as the time of the frame of the
observer (see \eqref{eq:galileanCoordTrans}). The coordinates
$x',y',z',t'$ define the marker's progressive position in acoustic
space-time, and $x_s',y_s',z_s',\tau'$ define the starting position of
the marker. Therefore we can define the 4-vectors $\xi$ and $\eta$
such that,
\begin{equation}
  \begin{aligned}
    \xi &= ct'\gamma_t' + x'\gamma_x' + y'\gamma_y' + z'\gamma_z', \\
    \eta &= c\tau'\gamma_t' + x_s'\gamma_x' + y_s'\gamma_y' + z_s'\gamma_z'.
  \end{aligned}
\end{equation}
Once again, $\eta$ is the single position in space-time at which the
marker on the wavefront starts, and $\xi$ is the collection of points
at which the marker can end up if it moves at the speed of sound in
any of the spatial directions.

We now consider the vector $\xi-\eta$, which points from the marker on
the wavefront's starting position to its current position, or in other
words the vector tangent to the marker's motion (this is analogous to
a ray \cite{Crighton:1992gm}). If we take the dot product of this vector
with itself we obtain (using \eqref{eq:spaceTimeSig}),
\begin{equation}
  (\xi-\eta)^2 = c^2(t'-\tau')^2 - (x'-x_s')^2 - (y'-y_s')^2 - (z'-z_s')^2.
\end{equation}
Invoking \eqref{eq:soundParticleMotion} we see that,
\begin{equation}
  (\xi-\eta)^2 = 0.
\end{equation}
It is important to note at this point that we are \emph{not} dealing
with a Euclidean space, and so this does not imply that $\xi-\eta$ is
a point. Instead it implies that $\xi-\eta$ is one of a set of vectors
referred to as null vectors.

To conclude, if a point on a wavefront is moving relative to the fluid
at the speed of sound, then the vector tangent to its motion in our
space time is a null vector, this is in fact independent of
frame. This is an important property of the acoustic space-time
interpretation, and we shall see it again in the following sections.

\subsection{Green's function for the convected wave equation}
\label{sec:GeneralSolnDerivation}
The Green's function for the convected wave equation satisfies,
\begin{equation}
  \label{eq:coodinateFreeGreensEq}
  \mathcal L G(\xi,\eta) = -\delta(\xi-\eta),
\end{equation}
where $\mathcal L$ is given by \eqref{eq:LinObserverFrame},
$\delta$ is a Dirac delta distribution, $\xi$ is a 4-vector giving the
position of the observer in acoustic space-time, and $\eta$ a 4-vector
giving the position of the source. We express these two 4-vectors in
the observer frame $\{\gamma_{\mu}\}$ and the Lorentzian frame
$\{\gamma_{\mu}\}''$ as,
\begin{equation}
  \begin{aligned}
    \xi &= ct\gamma_t + x\gamma_x + y\gamma_y + z\gamma_z =
    ct''\gamma_t'' + x''\gamma_x'' + y''\gamma_y'' + z''\gamma_z'', \\
    \eta &= c \tau \gamma_t + x_s\gamma_x + y_s\gamma_y + z_s\gamma_z =
    c \tau''\gamma_t'' + x_s''\gamma_x'' + y_s''\gamma_y'' + z_s''\gamma_z''.
  \end{aligned}
\end{equation}
The relations between the double prime and no prime coordinates are
given in \secref{sec:lorentzTransform}. If we write $\mathcal L$
in the $\{\gamma_\mu''\}$ frame (\eqref{eq:L_pp_coords}),
\eqref{eq:coodinateFreeGreensEq} reduces to the ordinary wave
equation for which the Green's function is well known and is given by
\cite[\S11.3]{Morse:1953vz},
\begin{equation}
  \label{eq:unconvectedGreensFunc}
  G(\xi,\eta) = \frac{\delta\left(t''-\tau'' - R''/c \right)}{4\pi R''},
\end{equation}
where
$R''\equiv\sqrt{(x''-x_s'')^2+(y''-y_s'')^2+(z''-z_s'')^2}$. Looking
carefully at the argument of the delta function, and recalling that
the $\{\gamma_\mu''\}$ frame must satisfy the space-time signature
(\eqref{eq:spaceTimeSig}), we see that in order for $G$ to be
non-zero, we must have,
\begin{equation}
  \label{eq:NullGFuncArg}
  (\xi-\eta)^2 = 0.
\end{equation}
Hence we see that the impulse is spreading out along the null
directions of our acoustic space-time, as was discussed in
\secref{sec:nullPropogation}. We can now express $G$ in the
observer-frame from \eqref{eq:nop_To_pp_coordTrans}, which yields,
\begin{equation}
  \label{eq:Rpp_to_nop}
  \begin{aligned}
    R'' &= \sqrt{(x-x_s)^2 + (y-y_s)^2 + (z-z_s)^2/\beta^2}, \\
    t'' - \tau'' &= \beta (t - \tau) + \frac{M}{\beta c}(z-z_s).
  \end{aligned}
\end{equation}
Substituting these results into \eqref{eq:unconvectedGreensFunc}
for $G$, we obtain an expression for $G$ in the observer
coordinates. We can use this to obtain a solution to
\eqref{eq:convectedWaveEq} for a general forcing $f(x,y,z,t)$,
\begin{equation}
  p = \int_{x_s}\int_{y_s}\int_{z_s}
      \frac{f\left(x_s,y_s,z_s,t + \frac{M}{\beta^2c}(z-z_s) -
      \frac{R''}{c}\right)}{4\pi\beta R''}
      \;dz_sdy_sdx_s,
\end{equation}
where we have used $R''$ to compress the notation, but we must
remember to evaluate it using \eqref{eq:Rpp_to_nop}. This is in
agreement with \cite{Chapman:2000ky}. However, it is more informative
to express $G$ in the fluid frame. Let us start by defining $R'$ as,
\begin{equation}
  \label{eq:RprimeDefn}
  R' \equiv \sqrt{(x'-x_s')^2+(y'-y_s')^2+(z'-z_s')^2},
\end{equation}
so it represents the distance between the source and the observer in
the fluid frame (note that the observer and source are moving in this
frame). Using \eqref{eq:NullGFuncArg}, and recalling that
$\{\gamma_\mu'\}$ satisfies the space-time signature, we see that when
$G$ is non-zero $R'=c(t'-\tau')$.

Since we only need to evaluate $G$ when the argument of the $\delta$
function is zero, using \eqref{eq:prime-to-double-prime}, we can
make the substitutions,
\begin{equation}
  \begin{aligned}
    \label{eq:Rpprime-Rprime}
    c(t''-\tau'') &= \beta c(t'-\tau')+ \frac{M}{\beta}(z'-z_s'), \\
    R'' &= \beta R'+ \frac{M}{\beta}(z'-z_s').
  \end{aligned}
\end{equation}
Substituting these results into \eqref{eq:unconvectedGreensFunc},
we obtain,
\begin{equation}
  \begin{aligned}
  G(\xi,\eta) = \frac{\delta\left(t'-\tau'-R'/c\right)}{4\pi \beta R''}
  \end{aligned}
\end{equation}
This corresponds to the classical form of the Green's function for the
convected wave equation \cite{Garrick:1953txa}. We can simplify it
further by expressing $R''$ in fluid coordinates~\cite{Hanson:1995ho}.
From \eqref{eq:lorentzian-Rpp} of \secref{sec:polar-coord-eucl},
${\beta R''=R'(1+M\cos\theta')}$, where $\theta'$ is the angle from
the flow to the vector $\xi - \eta$ in the fluid frame, we have
\begin{equation}
  G(\xi,\eta) = \frac{1}{1 + M\cos \theta'}
  \frac{\delta\left(t'-\tau' - R'/c \right)}{4\pi R'}.
\end{equation}
Thus, the Green's function in a uniform flow is identical to that in a
quiescent medium when expressed in the fluid frame, apart from a
convective amplification factor $1/(1+M\cos\theta')$. Sound is
amplified when the observer is upstream of the source and it is
reduced when the observer is downstream of the source, \new{as explained by
  \cite{Dowling:1983vo}.}

\subsection{\new{Visualising the source and observer positions on a
    plane}}
\label{sec:polar-coord-eucl}
It is convenient to \new{visualise} the position of the source and the
observer on a familiar Euclidean plane. This section details how this
can be achieved. It also relates the terminology introduced in this
paper with the widely used emission (or retarded) and reception
coordinates.

To simplify the analysis this section will use polar coordinates. In
the observer frame,
\begin{align}
  \label{eq:cylindrical}
  r &\equiv \sqrt{x^2 + y^2}, & \theta &\equiv \tan^{-1}(z/r),
  &\xi &= c t \gamma_t + r \gamma_r + z \gamma_z,
\end{align}
where $\gamma_r\equiv\cos\phi\gamma_x+\sin\phi\gamma_y$ and
$\tan\phi=y/x$. The observer spatial location can be pictured in the
$(\gamma_z, \gamma_r)$ plane.  We similarly define the polar
coordinates $(r', \theta')$ and $(r'', \theta'')$ in the fluid and
Lorentzian frames respectively.

Unfortunately, the plane $(\gamma_z'', \gamma_r'')$ is different from
the plane $(\gamma_z', \gamma_r')$ (which is the same as the
$(\gamma_z, \gamma_r)$ plane): the spatial coordinates in the
Lorentzian frame live in a separate plane from that of the observer
and fluid frames. To get around this difficulty, we simply map all the
polar coordinates to the same Euclidean 2D plane equipped with an
orthonormal frame $\{\bm e_z, \bm e_r\}$: the $\gamma_z$, $\gamma_z'$
and $\gamma_z''$ all map to $\bm e_z$, while $\gamma_r$, $\gamma_r'$
and $\gamma_r''$ all map to $\bm e_r$. Using this procedure, the
observer position is associated with the following Euclidean vectors:
\begin{align}
  \label{eq:euclidian-vectors}
  \bm x &\equiv z\bm e_z+r\bm e_r \\
  \bm x' &\equiv z'\bm e_z+r'\bm e_r' = (z-Mct)\bm e_z + r\bm e_r, \\
  \bm x'' &\equiv z''\bm e_z+r''\bm e_r''=(z/\beta)\bm e_z+r\bm e_r,
\end{align}
where we have expressed all the coordinates in terms of $r$, $z$ and
$t$ from \eqref{eq:galileanCoordTrans} and
\eqref{eq:nop_To_pp_coordTrans}.

Similarly, we define Euclidean source vectors as,
\begin{align}
  \label{eq:9}
  \bm x_s &\equiv z_s\bm e_z+r_s\bm e_r,\\
  \bm x_s' &\equiv z_s'\bm e_z+r_s'\bm e_r=(z_s-Mc\tau)\bm e_z+r_s\bm e_r,\\
  \bm x_s'' &\equiv z_s''\bm e_z+r_s''\bm e_r=(z_s/\beta)
  \bm e_z+r_s\bm e_r.
\end{align}
The source and observer Euclidean vectors are represented on
\figref{fig:sketch}. The figure helps \new{visualise} the relationships
between the source and observer positions in the familiar Euclidean
plane, as well as the role of the emission time $\tau$, reception time
$t$ and Mach number $M$.

If we add the point $\bm x_e$ along the $z$-axis to form the
parallelogram between $\bm x_s'$, $\bm x'$, $\bm x$ and $\bm x_e$
(\figref{fig:emission-coordinates}), we have,
\begin{align}
  \label{eq:2}
  \bm x_e &\equiv \bm x_s' + M c t \bm e_z = \bm x_s + M c (t - \tau)
  \bm e_z = \bm x_s + M R' \bm e_z,
\end{align}
which corresponds to the widely used ``emission'' (or retarded) source
position~\cite{Dowling:1983vo}. The emission source position is
sometimes called the virtual source position or the convected source
position~\cite{Sinayoko:2013iy}, since it can be obtained by convecting
the source $\bm x_s$ with the flow between emission time and reception
time. It is also common to define the ``reception'' source
position~\cite{Dowling:1983vo}, which corresponds to the source
position in the observer frame, i.e. $\bm x_s$.

The emission position is used to define \textit{emission coordinates}
$(R_e, \theta_e)$ (\figref{fig:emission-coordinates}), and by
construction we immediately see that $R_e\equiv|\bm x-\bm x_e|=R'$ and
$\theta_e=\theta'$. The emission position $\bm x_e$ is typically
regarded as the source position in the fluid frame. We now see that
this is not the case, since the source position in the fluid frame is
given by $\bm x_s'$. Furthermore, \figref{fig:emission-coordinates} is
typically used to derive $R_{e}$, but this requires solving a
quadratic equation. In contrast, \new{the expression for $R'$ derived in
  the previous section (see \eqref{eq:Rpprime-Rprime}) used} only
hyperbolic trigonometry and linear equations. This is a good example
of how acoustic space-time provides both physical insight and more
powerful algebraic tools.
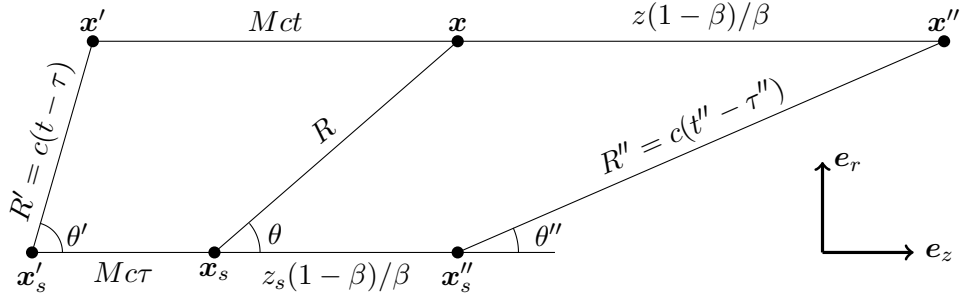
\begin{figure}[hbtp]
  \centering
  \begin{tikzpicture}[scale=0.8]
    \coordinate (xs) at (0,0);
    \coordinate (xsp) at (-3,0);
    \coordinate (xspp) at (4,0);
    \coordinate (xmxs) at (4,3.5);
    \coordinate (xpmxsp) at (1,3.5);
    \coordinate (xppmxspp) at (8,3.5);
    \draw[->,very thick] ($2.5*(xspp)$) -- ($2.5*(xspp)+(1.5,0)$)
    node[right] {$\eb_z$};
    \draw[->,very thick] ($2.5*(xspp)$) -- ($2.5*(xspp)+(0,1.5)$)
    node[right] {$\eb_r$};
    \draw (xsp) -- (xs) node[below,midway] {$Mc\tau$};
    \draw (xs) -- (xspp) node[below,midway] {$z_s(1-\beta)/\beta$};
    \draw (xspp) -- ($1.4*(xspp)$);
    \draw ($(xsp)+(xpmxsp)$) -- ($(xs)+(xmxs)$) node[above,midway] {$Mct$};
    \draw ($(xs)+(xmxs)$) -- ($(xspp)+(xppmxspp)$)
    node[above,midway] {$z(1-\beta)/\beta$};
    \draw (xs) -- ($(xs)+(xmxs)$) node[midway,rotate=40,yshift=8] {$R$};
    \draw (xsp) -- ($(xsp)+(xpmxsp)$)
    node[midway,rotate=75,yshift=8] {$R'=c(t-\tau)$};
    \draw (xspp) -- ($(xspp)+(xppmxspp)$)
    node[midway,rotate=24,yshift=8] {$R''=c(t''-\tau'')$};
    \fill (xs) circle[radius=0.1cm] node[below] {$\xb_s$};
    \fill (xsp) circle[radius=0.1cm] node[below] {$\xb_s'$};
    \fill (xspp) circle[radius=0.1cm] node[below] {$\xb_s''$};
    \fill ($(xs)+(xmxs)$) circle[radius=0.1cm] node[above] {$\xb$};
    \fill ($(xsp)+(xpmxsp)$) circle[radius=0.1cm] node[above] {$\xb'$};
    \fill ($(xspp)+(xppmxspp)$) circle[radius=0.1cm] node[above] {$\xb''$};
    \draw plot[domain=0:40] ({0.75*cos(\x)},{0.75*sin(\x)});
    \node at ($(xs)+(1,0.3)$) {$\theta$};
    \draw plot[domain=0:75] ({-3+0.5*cos(\x)},{0.5*sin(\x)});
    \node at ($(xsp)+(0.75,0.3)$) {$\theta'$};
    \draw plot[domain=0:24] ({4+cos(\x)},{sin(\x)});
    \node at ($(xspp)+(1.5,0.3)$) {$\theta''$};
  \end{tikzpicture}
  \caption{The source and observer locations can be represented in the
    same plane as $\bm x_{s}$ and $\bm x$ in the observer frame, $\bm
    x_{s}'$ and $\bm x'$ in the fluid frame, and $\bm x_{s}''$ and
    $\bm x''$ in the Lorentzian frame. The flow is uniform and from
    left to right.}
  \label{fig:sketch}
\end{figure}
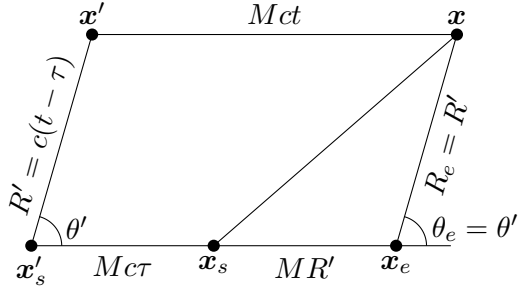
\begin{figure}[hbt]
  \centering
  \begin{tikzpicture}[scale=0.8]
    \coordinate (xs) at (0,0);
    \coordinate (xsp) at (-3,0);
    \coordinate (xmxs) at (4,3.5);
    \coordinate (xpmxsp) at (1,3.5);
    \coordinate (xe) at (3,0);
    \draw (xsp) -- (xs) node[midway,below] {$Mc\tau$};
    \draw (xs) -- (xe) node[midway,below] {$MR'$};
    \draw (xe) -- ($1.3*(xe)$);
    \draw ($(xsp)+(xpmxsp)$) -- ($(xs)+(xmxs)$) node[above,midway] {$Mct$};
    \draw (xs) -- ($(xs)+(xmxs)$);
    \draw (xsp) -- ($(xsp)+(xpmxsp)$)
    node[midway,rotate=75,yshift=8] {$R'=c(t-\tau)$};
    \draw (xe) -- ($(xs)+(xmxs)$)
    node[midway,rotate=75,yshift=-8] {$R_e=R'$};
    \fill (xs) circle[radius=0.1cm] node[below] {$\xb_s$};
    \fill (xsp) circle[radius=0.1cm] node[below] {$\xb_s'$};
    \fill ($(xs)+(xmxs)$) circle[radius=0.1cm] node[above] {$\xb$};
    \fill ($(xsp)+(xpmxsp)$) circle[radius=0.1cm] node[above] {$\xb'$};
    \fill (xe) circle[radius=0.1cm] node[below] {$\xb_e$};
    \draw plot[domain=0:75] ({-3+0.5*cos(\x)},{0.5*sin(\x)});
    \node at ($(xsp)+(0.75,0.3)$) {$\theta'$};
    \draw plot[domain=0:75] ({3+0.5*cos(\x)},{0.5*sin(\x)});
    \node at ($(xe)+(1.3,0.3)$) {$\theta_e=\theta'$};
  \end{tikzpicture}
  \caption{The emission coordinates $(R_{e}, \theta_{e})$, measured
    from the emission source position $\bm x_{e}$ (also called
    ``retarded'' or ``convected'' source position), correspond physically
    to the coordinates $(R', \theta')$ measured from the source
    position $\bm x_s'$ in the fluid frame. Indeed, the points $\bm
    x_s'$ (source in fluid frame), $\bm x_s$ (source position in
    observer frame), $\bm x$ (observer in observer frame) and $\bm x'$
    (observer in fluid frame) form a parallelogram.}
  \label{fig:emission-coordinates}
\end{figure}
Another example is that it is common to
introduce~\cite{Garrick:1953txa,Hanson:1995ho} the
amplitude radius $S=\beta R''$. However there is no clear physical
interpretation for the amplitude radius in terms of the source and
observer positions. We believe that $S$ is simply another way of
designating $R''$, which does have a clear physical interpretation as
the distance between the source and the observer in the Lorentzian
frame. We have therefore refrained from introducing $S$ in this paper.
\figref{fig:sketch} can also be used to derive relationships
between the spherical coordinates $(R,\theta)$, $(R',\theta')$ and
$(R'', \theta'')$. These relationships are provided in
\secref{sec:polar-coordinates}.

\subsection{Doppler shift}
\begin{figure}[!hbt]
  \centering
  \begin{tikzpicture}
    \coordinate (xo) at (5,2);
    \coordinate (Msf) at (1.5,2);
    \coordinate (Mof) at (1.5,-1.3);
    \fill (0,0) circle[radius=0.06cm];
    \fill (xo) circle[radius=0.06cm];
    \draw[dashed] (0,0)node[below] {$\xb_S'$}--(xo)node[above] {$\xb_O'$};
    \draw[ultra thick,->] ($0.3*(xo)$) -- ($0.4*(xo)$)
    node[below right] {$\kb'=(\omega'/c)\nb'$};
    \draw[thick,->] (0,0) -- (Msf) node[above] {$\Mb_{SF}$};
    \draw[thick,->] (xo) -- ($(xo)+(Mof)$) node[below] {$\Mb_{OF}$};
  \end{tikzpicture}
  \caption{Consider a sound wave travelling from source position $\bm
    x_S'(\tau)$ at emission time to observer position $\bm x_O'(t)$ at
    reception time in the fluid frame (the flow Mach number is 0). The
    source and observer are moving uniformly at Mach $\bm M_{SF}$
    and $\bm M_{OF}$ respectively. The wavenumber vector $\bm k'$
    can be expressed as $(\omega'/c) \bm n'$, where $\omega'$ is the
    wave frequency in the fluid frame and $\bm n'$ the unit vector
    from the source to the observer in the fluid frame.}
  \label{fig:doppler-problem}
\end{figure}
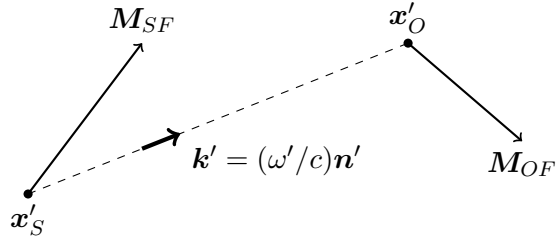
Consider an acoustic wave propagating in a fluid, relative to which a
source and observer are moving in uniform, rectilinear (but different)
motion. The terms ``source'' and ``observer'' are used to enable us to
differentiate between the two, but we could equally think of them as
two different observers.

In the Euclidean fluid frame, let $\bm x_O'$ denote the observer
position, $\bm x_S'$ the source position, $\bm M_{OF}$ the Mach
number vector of the observer relative to the fluid and $\bm M_{SF}$
the source Mach number vector. The problem is illustrated in
\figref{fig:doppler-problem}. We seek the Doppler shift between
the observer frequency $\omega_O$ and the source frequency $\omega_S$.

From \eqref{eq:freq_components_gppframe1}, the relationship
between the frequency in the observer frame $\omega_O$ and the
frequency in the fluid frame $\omega_F'$ can be expressed in vector
form as,
\begin{equation}
  \label{eq:omegaO}
  \omega_O = (\omega_F' - c\bm M_{OF}\cdot \bm k'),
\end{equation}
where $\bm k'$ is the wavenumber vector in Euclidean space. Using the
dispersion relation in the fluid frame, $|\bm k'|=\omega_F'/c$
\eqref{eq:omegaO} becomes,
\begin{equation}
  \label{eq:omegaO_rearranged}
  \omega_O = \omega_F' (1 - \bm M_{OF}\cdot \bm n'),
\end{equation}
where $\bm n'=\bm k'/|\bm k'|$ is the normalized wavenumber vector
in the fluid frame; $\bm n'$ is the unit vector from the source
position at emission time to the observer position at reception time
in the fluid frame, as illustrated in
\figref{fig:doppler-problem}.

Similarly, we can readily express the frequency in the source frame
$\omega_S$ as,
\begin{equation}
  \label{eq:omegaS}
  \omega_S = \omega_F'(1 - \bm M_{SF}\cdot \bm n').
\end{equation}

Taking the ratio of \eqref{eq:omegaO_rearranged} and
\eqref{eq:omegaS} yields,
\begin{equation*}
  \frac{\omega_O}{\omega_S} =
  \frac{ 1 - \bm M_{OF}\cdot \bm n'}
       { 1 - \bm M_{SF}\cdot \bm n'},
\end{equation*}
This is the classical result for the Doppler shift for a source moving
relative to a stationary observer embedded in a uniform flow
\cite{Sinayoko:2013iy,Rozenberg:2010ea}. It is interesting to note that we
have not used any partial derivative nor the concept of instantaneous
frequency \cite{Rienstra:2013vd} to arrive at this result.

\section{Geometric Algebra}
\label{sec:GeometricAlgebra}
\subsection{Introduction}
Geometric Algebra hinges around the introduction of a new ``geometric
product'' that allows us to multiply vectors and, in general,
multivectors. A geometric product is one that is associative,
distributive over addition, and such that $\xi^2 \in \mathbb R$. These
three properties provide a powerful algebraic structure to a space
equipped with the geometric product. For example, it becomes possible
to invert vectors and therefore to solve numerous equations
analytically. This avoids the need to introduce coordinates and to
re-write the equations in terms of matrices or tensors, which are more
cumbersome and obscure the geometric interpretation.

The geometric product of vectors can be used to define an inner and
outer product (denoted by $\cdot$ and $\wedge$) by taking the
symmetric and antisymmetric part respectively. Blades are
obtained by multiplying multiple vectors using the outer product. For
example, multiplying two vectors generates a bivector, which
corresponds to a directed plane element. Similarly, multiplying three
vectors yields a trivector, a directed volume element. This
generalizes to arbitrary (integer) dimensions. Multivectors are then
obtained from linear combinations of these blades.

The language of GA is sufficiently powerful to model most physical
phenomena from quantum mechanics to general relativity
\cite{Doran:2003jd,Hestenes:2002tv} and is regarded as a universal
mathematical language for physics and engineering
\cite{Lasenby:2000dh}. It has been applied to electromagnetics
\cite{Abbott:2010wy}, to solid mechanics \cite{McRobie:1999wv},
and to fluid mechanics \cite{Cibura:2008wz}.

From this point on, if two vectors appear adjacent to each other, a
geometric product is implied. For a thorough introduction to the
basics of the algebra see \cite[\S4]{Doran:2003jd}, for information
regarding space-time see \cite[\S5]{Doran:2003jd}, and for information
regarding calculus in this new algebra see \cite[\S6]{Doran:2003jd}.

\subsection{The Lorentz Transform}
\label{sec:GALorentzTrans}
Lorentz transforms have a very neat interpretation in geometric
algebra, which will now be illustrated using the above example. We
repeat the definitions of the Lorentzian frame $\{\gamma_\mu''\}$,
\begin{equation}
  \gamma_x'' = \gamma_x',\quad \gamma_y'' = \gamma_y',\quad
  \gamma_z'' = \frac{1}{\beta}\left(\gamma_z' -
  M\gamma_t'\right),\quad
  \gamma_t'' = \frac{1}{\beta}\left(\gamma_t' - M\gamma_z'\right).
\end{equation}
We now introduce the `hyperbolic angle' $\alpha$,
\begin{equation}
  \tanh\alpha \equiv M,
\end{equation}
so that $\cosh\alpha=1/\sqrt{1-\tanh^2\alpha}=1/\beta$.

The $\gamma_t''$ vector now becomes,
\begin{equation}
\label{eq:hrot1}
  \begin{aligned}
    \gamma_t'' &= \cosh\alpha\gamma_t' - \sinh\alpha\gamma_z'
    = \left(\cosh\alpha - \sinh\alpha\gamma_z'\gamma_t'\right) \gamma_t'
    = \exp\left(-\alpha\gamma_z'\gamma_t'\right)\gamma_t',
  \end{aligned}
\end{equation}
where the final equality follows from the power series expansions of
$\exp$, $\sinh$ and $\cosh$. Similarly, $\gamma_z''$ can be expressed
as,
\begin{equation}
\label{eq:hrot2}
  \begin{aligned}
    \gamma_z'' = \cosh\alpha\gamma_z' - \sinh\alpha\gamma_t' =
    \exp\left(-\alpha\gamma_z'\gamma_t'\right)\gamma_z'.
  \end{aligned}
\end{equation}

There is a very compact way of expressing rotations in geometric
algebra by using \textit{rotors}:
\begin{equation}
  \gamma_t'' = R\gamma_t'\tilde R,\quad
  \gamma_z'' = R\gamma_z'\tilde R,
\end{equation}
where we have introduced the rotor,
\begin{equation}
  R = \exp\left(-\frac{\alpha\gamma_z'\gamma_t'}{2}\right),\quad
  \tilde R = \exp\left(\frac{\alpha\gamma_z'\gamma_t'}{2}\right),
\end{equation}
where the over tilde denotes reversion~\cite{Doran:2003jd}. We note that
$R\tilde R=1$, and so we can write,
\begin{equation}
  \gamma_x'' = \gamma_x' = R\tilde R\gamma_x' =
  R\gamma_x'\tilde R,
\end{equation}
since $\gamma_x'$ commutes with $\gamma_z'\gamma_t'$. The
same can be said for $\gamma_y''$, and so we are able to write the
whole Lorentz transformation as,
\begin{equation}
\label{eq:hyperbolic_rotation}
  \gamma_\mu'' = R\gamma_\mu'\tilde R,\quad R =
  \exp\left(-\frac{\alpha\gamma_z'\gamma_t'}{2}\right).
\end{equation}
The rotor $R$ is a general way of encoding transformations that
preserve lengths and angles, as defined by the inner product. If a
rotor is constructed in a Euclidean space, it encodes a rotation
(hence the name), but since our space is hyperbolic, the example above
is a Lorentz transformation.

If our space was the space-time used in relativity (i.e. if $c$ was
the speed of light), the Lorentz transformation would represent the
transformation between the two frames constructed by two observers,
with the $\{\gamma_\mu''\}$ observer travelling at speed $Mc$
backwards along the $z$ axis of the $\{\gamma_\mu'\}$ observer.

If we interpret this back in our acoustic space-time, we might say
that the Lorentz transformation moves us from the fluid frame, back to
a stationary frame (moving with the observer), but ensuring that our
stationary frame still satisfies the space-time signature, which our
original stationary frame $\{\gamma_\mu\}$ did not.

\subsection{Geometric interpretation of the coordinate transformations}
\begin{figure}[t]
  \centering
  \includegraphics[width=\textwidth]{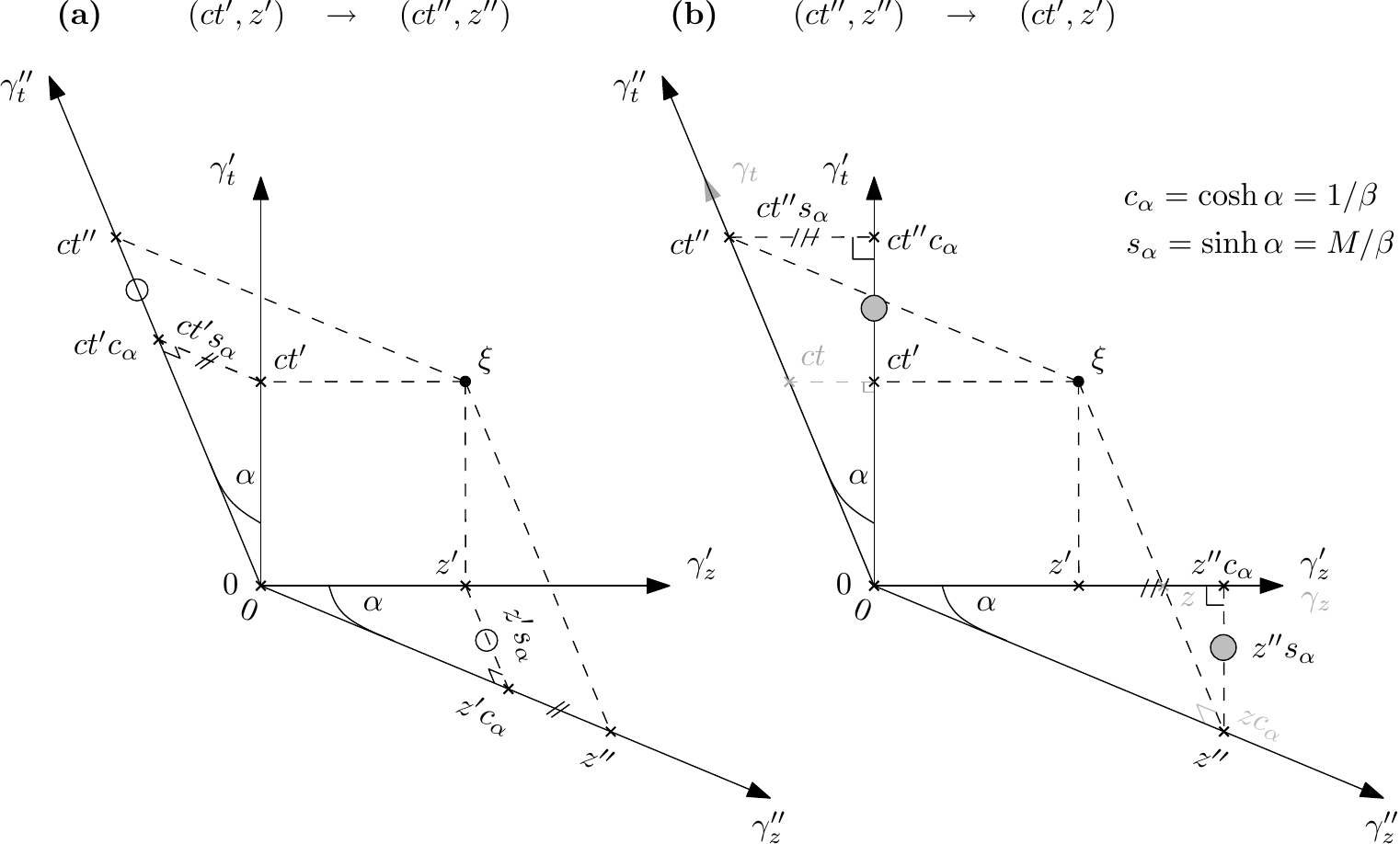}
  \caption{Geometric interpretation of the transformations between the
    fluid frame $\{\gamma'_{\mu}\}$ and the Lorentzian frame
    $\{\gamma''_{\mu}\}$: (a) transformations from fluid coordinates
    $(ct', z')$ to Lorentzian coordinates $(ct'', z'')$; (b)
    transformations from Lorentzian coordinates $(ct'', z'')$ to fluid
    coordinates $(ct', z')$ and observer coordinates $(ct, z)$. The
    geometry of the fluid frame is Euclidean but the geometry of the
    Lorentzian frame is hyperbolic. In particular the angle $\alpha =
    \tanh^{-1} M$ is a hyperbolic angle. These two figures can be used
    to recover in a geometrical way the coordinate transformations
    between the observer frame, the fluid frame and the Lorentzian
    frame.}
\label{fig:coordinate-transformations}
\end{figure}

\eqref{eq:hrot1} and \eqref{eq:hrot2} provide a simple
geometric interpretation of the transformation: the Lorentzian frame
$\{\gamma''_{\mu}\}$ can be obtained from the fluid frame
$\{\gamma'_{\mu}\}$ by means of a hyperbolic rotation of angle
$\alpha$. This is illustrated in
\figref{fig:coordinate-transformations}. Using the rules of
hyperbolic trigonometry, it is possible to recover the coordinate
transformations of \eqref{eq:galileanCoordTrans},
\eqref{eq:prime-to-double-prime} and \eqref{eq:nop_To_pp_coordTrans}
directly from \figref{fig:coordinate-transformations}.

For example using the hyperbolic right angle triangles illustrated in
\figref{fig:coordinate-transformations}, we have,
\begin{align}
  \label{eq:double-prime-to-prime-geometrically}
  ct'' = ct' \cosh \alpha + z' \sinh \alpha = (1/\beta)(ct' + M z') \\
  z'' = z' \cosh \alpha + ct' \sinh \alpha = (1/\beta)(z' + M c t'),
\end{align}
which corresponds to \eqref{eq:prime-to-double-prime}. Note that
the right angle triangles look different from the usual Euclidean
ones: the hypotenuse is shorter than the two other sides of the
triangle, which is typical of hyperbolic geometry.

Similarly, one can invert the above transformations by projecting the
Lorentzian coordinates onto the fluid frame, as illustrated in
\figref{fig:coordinate-transformations},
\begin{align}
  ct' = ct'' \cosh \alpha - z'' \sinh \alpha = (1/\beta)(ct'' - M z'') \\
  z' = z'' \cosh \alpha - ct'' \sinh \alpha = (1/\beta)(z'' - M c t'').
\end{align}
Using the same figure, one can readily express the Lorentzian
coordinates in terms of the observer coordinates,
\begin{align}
  \label{eq:no-prime-to-double-prime-geometrically}
  &z'' = z \cosh \alpha = z/\beta, \\ &ct' = (ct / \beta) \cosh \alpha =
  ct, \quad \text{so from
    \eqref{eq:double-prime-to-prime-geometrically}} \quad ct'' =
  \beta ct + M z/\beta
\end{align}
which corresponds to \eqref{eq:nop_To_pp_coordTrans}.

Thus, the Lorentzian frame in which the wave equation retains its
simplest form obeys the laws of hyperbolic geometry. This does not
appear to have been fully appreciated in the aeroacoustics literature
and may simplify future analyses.

\subsection{The propagation operator $\mathcal L$ and the Laplacian in acoustic space-time}
\label{sec:GAVectorDerivative}
If we have an arbitrary frame $\{ e_i\}$ $i=1,\hdots,n$ with
reciprocal frame $\{ e^i\}$, an arbitrary vector $\xi$ may
be written,
\begin{equation}
  \xi = x^i e_i \text{ where } x^i = \xi\cdot e^i,
\end{equation}
where the Einstein summation convention for repeated indices has been
employed. For an arbitrary field $F(\xi)$, the vector derivative
$\nabla$ is defined such that, for any vector $a$,
\begin{equation}
   a\cdot\nabla F(\xi) = \lim_{\epsilon\rightarrow0}
  \frac{F(\xi+\epsilon a) - F(\xi)}{\epsilon},
\end{equation}
where we assume that the limit exists and is well defined. Note that
$F$ can in general be any multivector field. It can be shown that the
vector derivative can be written,
\begin{equation}
  \label{eq:vectorDerivativeDefn}
  \nabla = \sum_{i=1}^n e^i\frac{\partial}{\partial x^i}
  = \sum_{i=1}^n e^i\partial_{x^i}.
\end{equation}
This defines the shorthand $\partial_x$ for a partial derivative with
respect to $x$. Note that $\nabla$ is frame independent.

Now let us consider the operator $\nabla\cdot\nabla$ in the acoustic
space-time frames defined above. Using the definition in
\eqref{eq:vectorDerivativeDefn} $\nabla$ is given by,
\begin{equation}
  \begin{aligned}
    \nabla &= \gamma^t\partial_{ct} + \gamma^x\partial_x +
    \gamma^y\partial_y + \gamma^z\partial_z \\
    &= {\gamma^t}'\partial_{ct'} + {\gamma^x}'\partial_{x'} +
    {\gamma^y}'\partial_{y'} + {\gamma^z}'\partial_{z'} \\
    &= {\gamma^t}''\partial_{ct''} + {\gamma^x}''\partial_{x''} +
    {\gamma^y}''\partial_{y''} + {\gamma^z}''\partial_{z''}.
  \end{aligned}
\end{equation}
In all the frames defined the basis vectors do not vary over the
space. Using this and the definitions of the basis vectors given
above, we can show that,
\begin{equation}
  \begin{aligned}
  \nabla\cdot\nabla &= - \partial_x\partial_x - \partial_y\partial_y
  - \partial_z\partial_z + (\partial_{ct}+M\partial_z)^2 \\
  &= - \partial_{x'}\partial_{x'} - \partial_{y'}\partial_{y'}
  - \partial_{z'}\partial_{z'} + \partial_{ct'}\partial_{ct'} \\
  &= - \partial_{x''}\partial_{x''} - \partial_{y''}\partial_{y''}
  - \partial_{z''}\partial_{z''} + \partial_{ct''}\partial_{ct''}.
  \end{aligned}
\end{equation}
Hence it becomes clear that,
\begin{equation}
  \mathcal L = -\nabla\cdot\nabla.
\end{equation}

Our wave equation may therefore be written,
\begin{equation}
  \nabla\cdot\nabla p(\xi) = f(\xi),
\end{equation}
and depending on which frame we expand $\nabla\cdot\nabla$ in, we
obtain either the convected or the un-convected wave equation. It is
now clear why the Lorentz transform does not alter the form of
$\mathcal L$. When the Lorentz transform acts on a frame that
satisfies the \new{simple} space-time \new{metric}
(\eqref{eq:spaceTimeSig}), the resulting frame also satisfies this
\new{metric}, hence the expansion of $\mathcal L = -\nabla\cdot\nabla$
is not altered. However, when a Galilean transform is used, the
resulting frame is no longer orthogonal, and this is why the expansion
of $\nabla\cdot\nabla$ is more complex.

\section{Conclusions}

This paper presents a geometric framework to interpret the
transformations used to solve the equation for sound propagation in
uniform flow. By writing the wave operator in terms of the Laplacian
of the acoustic space-time we are able to show clearly why the wave
operator takes complex and simple forms in the different frames.  The
framework allows the transformation to be understood in terms of a
combined Galilean and Lorentz transform.  The Galilean transform
simplifies the wave operator but moves us to a frame in motion with
respect to the observer. The Lorentz transform takes us back to a
frame that is stationary with respect to the observer, while keeping
the wave operator simple. The use of Geometric algebra makes it clear
that the Lorentz transform is really a generalised rotation.

The transformation derived in this way is used to give simple
derivations of Green's functions in the different frames. The geometric
framework allows sound propagation to be understood in terms of the
null vectors of the space, which are not dependent on the frame. A
simple interpretation for transformations in frequency-wavenumber
space in terms of the reciprocal frame is also provided, which is
used to derive expressions for Doppler shifts.

We hope that this investigation will pave the way to more complex
transformations applicable when the background flow is non-uniform.
Here we have used a flat space-time to understand uniform flow
transformation, and the natural extension would be to consider a
curved acoustic space-time.



\appendix

\section{The Reciprocal Frame}
\label{app:reciprocalFrames}
Here we follow \cite{Doran:2003jd}. Let us assume that we have a set
of $n$ linearly independent vectors $\{\bm e_k\}$ that span an
arbitrary $n$ dimensional space (arbitrary means that the frame can be
Euclidean or hyperbolic, or indeed can have any \new{metric}). We refer
to this set of $n$ vectors as our \emph{frame}. Associated with any
frame is a reciprocal frame of $n$ vectors, $\{\bm e^k\}$, defined by
the property,
\begin{equation}
  \label{eq:app_reciprocalRequirement}
  \bm e^i\cdot\bm e_j = \delta^i_j,\quad \forall i,j=1,\hdots,n
  \qquad \delta^i_j = \left\{\begin{array}{ll}
    1 & i=j \\ 0 & i\neq j
  \end{array}\right. .
\end{equation}

\subsection{Vector Components}
The basis vectors $\{\bm e_k\}$ are linearly independent, so any
vector $\bm a$ can be written uniquely in terms of this set as,
\begin{equation}
  \label{eq:vectorComponents}
  \bm a =  \sum_{i=1}^n a^i\bm e_i = \sum_{i=1}^na_i\bm e^i.
\end{equation}
The set of scalars $(a^1,\hdots,a^n)$ are the components of the vector
$\bm a$ in the $\{\bm e_k\}$ frame. To find these components we form,
\begin{equation}
  \bm a\cdot\bm e^i = a^j\bm e_j\cdot\bm e^i =
  a^j\delta^j_i = a^i,
\end{equation}
\begin{equation}
  \bm a\cdot\bm e_i = a_j\bm e^j\cdot\bm e_i =
  a_j\delta^i_j = a_i.
\end{equation}
This derivation of how to find components explains the use of sub- and
superscripts in \eqref{eq:vectorComponents}. Note that if the
frame is orthonormal, the frame and its reciprocal are equivalent.

\subsection{Finding the Reciprocal Frame}
Given that the basis vectors $\{\bm e_k\}$ are linearly independent,
it must be possible to decompose every reciprocal vector as,
\begin{equation}
  \bm e^k = a^k_1\bm e_1 + \hdots + a^k_n\bm e_n,
\end{equation}
for some set of coefficients $a^k_i$. Given this and the requirement
in \eqref{eq:app_reciprocalRequirement}, we can obtain a system of
$n^2$ equations for the $n^2$ coefficients $a^k_i$. For large $n$
however, this will become rather cumbersome, and so we also present a
result from geometric algebra that gives us an explicit expression for
the reciprocal frame, which can be easily implemented in any symbolic
algebra package.

We define the volume element for the $\{\bm e_k\}$ frame, $E_n$, as,
\begin{equation}
  E_n = \bm e_1\wedge\bm e_2\wedge\cdots\wedge\bm e_n.
\end{equation}
Note that $E_n$ is a multiple of the pseudoscalar $I$. We can
construct the reciprocal frame as,
\begin{equation}
  \label{eq:findingRecipFrame}
  \bm e^j = (-1)^{j-1}\bm e_1\wedge\cdots\wedge\check{\bm e}_j
  \wedge\cdots\wedge\bm e_n \;E_n^{-1},
\end{equation}
where the check on $\check{\bm e}_j$ denotes that this vector is
excluded from the expression. We now justify this definition. $\bm
e^j$ must be perpendicular to all the $\{\bm e_i,i\neq j\}$, and hence
we form the exterior product of these $n-1$ vectors, and then project
onto the vector perpendicular to this $(n-1)$ volume element by
multiplying by $E_n^{-1}$. The fact that we used $E_n^{-1}$ rather
than $I$ ensures the correct normalisation. To prove this more
formally we make use of the property of the pseudoscalar $I$, that for
any vector $\bm a$ and multivector $A_r$ (of grade $r$)
\cite{Doran:2003jd},
\begin{equation}
  \bm a\cdot(A_rI) = (\bm a\wedge A_r) I.
\end{equation}

Let us define the scalar $C$ such that,
\begin{equation}
  E_n = C I.
\end{equation}
It follows that,
\begin{equation}
  E_n^{-1} = \frac{1}{C}I^{-1},
\end{equation}
and hence we can write,
\begin{equation}
  \begin{aligned}
    \bm e_i\cdot\bm e^j &= \frac{(-1)^{j-1}}{C}\bm e_i\cdot\left(
    (\bm e_1\wedge\cdots\wedge\check{\bm e}_j\wedge\cdots\wedge
    \bm e_n)I^{-1}\right) \\
    &= \frac{(-1)^{j-1}}{C}(\bm e_i\wedge\bm e_1\wedge\cdots
    \wedge\check{\bm e}_j\wedge\cdots\wedge\bm e_n)I^{-1}.
  \end{aligned}
\end{equation}
If $i\neq j$ then this contains $\bm e_i\wedge\bm e_i$ and hence is
zero. If $i=j$ we can write this as,
\begin{equation}
  \begin{aligned}
    \bm e_i\cdot\bm e^j &= \frac{1}{C}(\bm e_1\wedge\cdots
    \wedge\bm e_j\wedge\cdots\wedge\bm e_n) I^{-1} \\
    &= \frac{1}{C}(CI)I^{-1} = 1 \text{ as required.}
  \end{aligned}
\end{equation}

\section{Spherical coordinates}
\label{sec:polar-coordinates}
\begin{itemize}
\item Observer coordinates $(R,\theta)$:
\begin{align}
  \label{eq:observer-R}
  R &=  R' \sqrt{1 + M^{2} + 2 M \cos \theta'} = R''
  \sqrt{1 - M^{2}\cos^2\theta''} \\
  \cos \theta &= \frac{\cos\theta' + M}
       {\sqrt{1 + M^{2} + 2 M \cos \theta'}} =
       \frac{\beta \cos\theta''}{\sqrt{1 - M^2 \cos^2 \theta''}},\\
  \sin \theta &=  \frac{\sin \theta'}{\sqrt{1 + M^{2} + 2 M \cos \theta'}}
  = \frac{\sin\theta''}{\sqrt{1 - M^2 \cos^2 \theta''}};
\end{align}
\item Lorentzian coordinates $(R'',\theta'')$:
\begin{align}
  \label{eq:lorentzian-Rpp}
  R'' &=  \frac{R}{\beta} \sqrt{1 - M^{2}\sin^2\theta} =
  \frac{R'}{\beta} ( 1 + M \cos\theta'),\\
  \cos\theta'' &= \frac{\cos\theta}{\sqrt{1 - M^2\sin^2\theta}} =
  \frac{M + \cos \theta'}{ 1 + M \cos\theta'}, \\
  \sin \theta'' &= \frac{\beta\sin\theta}{\sqrt{1-M^2\sin^{2} \theta}}
  = \frac{\beta \sin\theta'}{1 + M \cos\theta'};
  \end{align}
\item Fluid coordinates $(R',\theta')$:
\begin{align}
  \label{eq:fluid-Rp}
  R' &=  \frac{R}{\beta^2} \left(\sqrt{1 - M^{2}\sin^2\theta} -
  M \cos \theta\right) = \frac{R''}{\beta} \left(1 - M \cos \theta''\right), \\
  \cos\theta' &= \frac{\cos \theta - M\sqrt{1 - M^2\sin^2\theta}}
              {\sqrt{1 - M^2\sin^2\theta} - M \cos\theta} =
              \frac{M - \cos \theta''}{M\cos\theta'' - 1}, \\
  \sin \theta' &= \frac{\beta^2 \sin\theta}{\sqrt{1 - M^2\sin^2\theta}
    - M \cos\theta} = \frac{\beta \sin \theta''}{1 - M \cos \theta''}.
\end{align}
\end{itemize}
It is interesting to note that the relationships involving the
Lorentzian coordinates are more compact than those between the
observer and fluid coordinates.

\bibliographystyle{rspub_unsrt}
\bibliography{GAbib}
\end{document}